\documentstyle[12pt,aasms4]{article}

\def\gappr{\mathrel{\vcenter{\offinterlineskip \hbox{$>$}
    \kern 0.3ex \hbox{$\sim$}}}}
\def\lappr{\mathrel{\vcenter{\offinterlineskip \hbox{$<$}
    \kern 0.3ex \hbox{$\sim$}}}}

\received{}
\revised{}
\singlespace

\begin{document}

\title{A Possible Mechanism for Wiggling Protostellar Jets  
From 3D Simulations in a Stratified Ambient Medium}

\author{Elisabete M. de Gouveia Dal Pino$^{1}$, Mark Birkinshaw$^{2,3}$, and 
Willy Benz$^{4}$}

\affil{$^1$ Instituto Astron\^{o}mico e Geof\'{\i}sico, Universidade 
de S\~{a}o Paulo, \\  
 Av. Miguel St\'efano 4200,  
(04301-904) S\~{a}o Paulo - SP, BRAZIL \\
E-mail: dalpino@astro1.iagusp.usp.br}

\singlespace
\affil{$^{2}$ 
Center for Astrophysics, 60 Garden St, Cambridge, MA, 02178, USA
}

\singlespace
\affil{$^{3}$ 
 Present Address: University of Bristol, Department of Physics, Bristol, UK
}

\singlespace
\affil{$^{4}$ 
University of Arizona, Steward Obs. and Lunar and Plan. Lab.,
Tuscon, AZ 85721, USA}

\vskip 2.0 cm
\hskip 1.5 cm {\bf To appear in the Astrophys. J. (Letters)}

\begin{abstract}
Most collimated supersonic protostellar jets show a collimated
wiggling, and knotty structure (e.g., the Haro 6-5B jet) and
frequently reveal a long gap between this structure and the terminal
bow shock. In a few cases, there is no evidence of such a terminal
feature.  We present three-dimensional smoothed particle
hydrodynamical simulations which suggest that this morphology may be
due to the interaction of the propagating cooling jet with a
non-homogeneous ambient medium.  In regions where the ambient gas has
an increasing density (and pressure) gradient, we find that it tends
to compress the cold, low-pressure cocoon of shocked material that
surrounds the beam, destroy the bow shock-like structure at the head,
and enhance beam focusing, wiggling, and internal traveling shocks. In
ambient regions of decreasing density (and pressure), the flow widens
and relaxes, becoming very faint. This could explain ``invisible''
segments in systems like the Haro 6-5B jet. The bow shock in these
cases could be a relic of an earlier outflow episode, as previously
suggested, or the place where the jet reappears after striking a
denser portion of the ambient medium.
\end{abstract}

\keywords{Hydrodynamics - stars: pre-main-sequence stars: 
mass loss - ISM: jets and outflows}

\clearpage

\section{Introduction}
Protostellar jets propagate into a complex ambient medium.  They are
immersed in molecular clouds which are generally inhomogeneous and
formed of many small, dense clouds which are sites of star formation
(e.g., Bally and Devine 1994).  Previous three- dimensional (3-D)
numerical modeling of continuous and intermittent jets emanating from
protostars were performed with beams propagating into homogeneous
ambient media (e.g., Gouveia Dal Pino and Benz 1993, 1994, hereafter
GB93, GB94; Chernin et al. 1994; hereafter CMGB; Stone and Norman
1994). These simulations have reproduced the basic features typically
observed in protostellar jets: the chain of more or less regularly
spaced knots along the beam, produced mainly in jets with intermittent
ejection; or the formation of bow shock-like structures at the jet
head, with a dense shell composed of filaments and blobs developed
from Rayleigh-Taylor and global thermal instabilities.  We here report
3-D numerical results for jets propagating through ambient gas with
power-law density (and pressure) distributions, and examine their
effects on the structure of the jets.

In recent work, concerned with environmental effects on extragalactic
jets, Wiita, Rosen and Norman (1990) and Hardee et al. (1992) have
performed 2-D hydrodynamical simulations of adiabatic, light jets
propagating into a stratified intergalactic medium. In contrast to
such extragalactic flows, the cooling distance of post-shock
protostellar jet gas is smaller than the jet radius and hence the
assumption of an adiabatic gas is inappropriate.  We here analyse the
effects of a stratified ambient medium on radiatively cooling,
initially heavy (denser than their surroundings) jets - a picture
generally believed to be more appropriate to protostellar jets (e.g.,
Mundt et al. 1990). A more detailed description of the results of this
investigation, with simulations covering a more extensive range of
parameters, is presented elsewhere (Gouveia Dal Pino and Birkinshaw
1995, hereafter GB95).

\section{ Numerical Method}
We solve the hydrodynamics equations using a modified version of the
three-dimensional Cartesian, gridless, Lagrangian smoothed particle
hydrodynamics (SPH) code described in GB93, GB94, CMGB, and GB95.  The
ambient gas is represented by a 3-D rectangular box filled with
particles. A collimated, supersonic jet of radius $R_j$ is
continuously injected into the bottom of the box, which has dimensions
of $\simeq 20 R_j$ in the x-axis and 12 $R_j$ in the transverse
directions (y and z). The boundaries of the box are assumed to be
continuous. The SPH particles are smoothed by a spherically symmetric
kernel function of width $h$. The initial values of $h$ were chosen to
be $0.5 R_j$ and $0.25 R_j$ for the ambient and jet particles,
respectively. The jet and the ambient gas are treated as a single
ionized fluid with an ideal gas equation of state and a ratio of
specific heats $\gamma = 5/3$. Radiative cooling (due to collisional
excitation and recombination) is implicitly calculated using the
local, time- independent cooling function for a gas of cosmic
abundances cooling from $T \simeq 10^6$ to $10^4$ K (Katz 1989).

We assume an initial isothermal ambient medium ($T_a=10^4$ K) with
density (and pressure) distribution stratified along the jet axis,
\begin{equation}
n_a(x) \, = \, n_a(x_o) [ \alpha \, (x - x_o) + 1 ]^{\beta},
\end{equation}
\noindent
where $n_a$ is the ambient number density, $x_o$ is the value of $x$
at the jet inlet (in units of the jet radius $R_j$), $\alpha$ is a
parameter of the model and the exponent $\beta \, = \, \pm \, 5/3$ for
positive and negative density gradients, respectively.  Such profiles
are consistent with the observed density distributions of clouds which
involve protostars (e.g., Fuller and Myers 1992). A negative density
gradient may represent, for example, the atmosphere that a jet
encounters as it propagates through a portion of the cloud that
envelops the parent source.  A positive density gradient will occur
when a jet enters an external cloud.  In the simulations the
atmosphere was held steady by an appropriate gravitational potential.

A supersonic jet propagating into a stationary ambient gas will
develop a shock pattern at its head, where the impacted ambient
material is accelerated by a forward bow shock and the jet material is
decelerated at the jet shock. Our models of hydrodynamical jets are
characterized by dimensionless parameters $\eta \, = \,
n_j(x_o)/n_a(x_o)$, the ratio between the jet and ambient number
densities at $x_o$; $M_a= v_j/c_a$ the ambient Mach number, where
$v_j$ is the jet injection velocity and $c_a$ is the sound speed in
the ambient material; and $q_{bs} = d_{cool}/R_j$, the ratio of the
cooling length in the postshock gas behind the bow shock to the jet
radius (which, for shock velocities $ > 90$ km/s is related to the
ratio measured in the postshock gas behind the jet shock, $q_{js}$,
via $q_{js} \simeq \, q_{bs} \eta^{-3}$ (cf., GB93). The parameters of
the simulations were chosen to resemble the conditions found in
protostellar jets.  Based on observations, we adopt the values $\eta =
1-10$; $v_j \simeq 400$ km/s; $M_a = 24$, and $R_j = 2 \times 10^{15}$
cm. We assume $\alpha = 0.5$ and $\beta = \pm 5/3$ in eq. 1. (See GB95
for a wider range of parameters.)

\section{ The Simulations}
Fig. 1 depicts the central density contours of three distinct jets
when they have propagated a distance $\simeq 20 R_j$. All flows were
initially in pressure equilibrium at the jet inlet, with $\eta (x_o)
\, = \, 3$, $n_a (x_o) =$ 200 cm$^{-3}$, $R_j = 2 \times 10^{15}$ cm,
$v_j = 398$ km s$^{-1}$, $M_a = 24$, $q_{bs} \simeq 8.1$, $q_{js}
\simeq 0.3$, and $R_j/c_a = 38.2$ yrs. In a) the ambient medium has
positive density gradient ($\beta =5/3$, $\alpha = 0.5$), in b) it is
homogeneous, and in c) it has negative density gradient ($\beta =
-5/3$, $\alpha = 0.5$). The cooling length parameters imply that the
ambient shocked gas is initially almost adiabatic whereas the shocked
jet material is highly radiative (e.g., Blondin, Fryxell and K\"onigl
1990, hereafter BFK; GB93).  The jets reach the end of the computation
domain at (in units of $R_j/c_a$): a) t= 1.65, b) t=1.15, and c)
t=0.95.  In the homogeneous medium case (Fig.  1b), we can identify
the same basic features detected in previous work (e.g., GB93, GB94,
CMGB).  The working surface, at the head of the jet, develops a dense
shell (with density $n_{sh} \sim \, 469 n_a$) of shocked jet material
which becomes Rayleigh-Taylor unstable (GB93) and breaks into blobs. A
low density cocoon containing post-shock jet gas is deposited around
the jet beam.  A dense shroud of shocked ambient gas envelops both the
beam and the cocoon.

In the increasing density medium (Fig. 1a), the cocoon is compressed
and pushed backwards by the increasing ram pressure of the ambient
gas.  The bow shock shape seen at the head of the jet propagating into
a uniform medium (Fig. 1b), is replaced in Fig. 1a by an elongated
structure, and the beam and shell ($n_{sh}/n_a (x_o) \simeq 783$) are
highly focused. The high pressure of the surrounding medium promotes
Kelvin-Helmholtz instabilities (e.g.  Birkinshaw 1991) with the
excitation of pinch and helical modes which collimate the beam, drive
internal shocks, and cause some jet twisting and flapping.  A similar
structure has been found for adiabatic jets propagating into a
homogeneous ambient medium (e.g., GB93), but there the collimation of
the jet is a result of the hot, high-pressure, post-shock jet gas in
the cocoon immediately behind the working surface (cf., BFK, GB93).
The internal shocks in the beam have a typical separation of few jet
diameters, in agreement with observations of internal knots in
protostellar jets.

In the decreasing density medium (Fig. 1c) the cocoon that surrounds
the beam becomes broad and relaxed, the jet shows no internal shocks,
and we clearly see an acceleration of the jet as it propagates due to
the decreasing ambient density and ram pressure.  Compared to the jet
propagating into the homogeneous medium (Fig. 1b), the beam is much
less collimated close to the head. An analysis of the density and
velocity maps indicates a ratio of the jet to head radius
$\epsilon^{1/2} \, \equiv \, R_j/R_h \simeq 0.8$.  This decollimation
is significantly larger for lower-density jets (GB95). The cold shell
at the working surface is thin, does not fragment, and has a smaller
overdensity ($n_{sh}/n_a (x_o) \simeq 33$) than the jet propagating
into the homogeneous ambient medium, which indicates the faintness of
radiative shocks at the head.  For an ambient medium with negative
density gradient, the cooling length behind the bow shock ($q_{bs}$)
should increase as the jet propagates downstream and the cooling length
behind the jet shock should decrease (e.g., GB93, GB95).  Thus the
bow-shock becomes more adiabatic as the beam propagates, while the jet
shock becomes more radiative but fainter since the jet shock velocity,
$v_{js} \simeq \, v_j - v_{ws}$, decreases as the bow shock speed
$v_{ws}$ increases.  The opposite is true in the case of the ambient
medium with positive density gradient.

Fig. 2 depicts jets with different values of $\eta$ propagating into
an $\alpha \, = 0.5$ ambient medium with positive ($\beta = 5/3$)
density (and pressure) stratification.  The jets reach the end of the
compution domain at (in units of $R_j/c_a$): a) t= 1.85, b) t=1.65,
and c) t=1.25. The highest-density jet reaches the end of the domain
earlier and is least affected by the increasing ambient density and
pressure gradients.

\section{Discussion and Conclusions}
What do these results have to do with the observed protostellar jets?
Our simulations show that jets propagating into portions of an ambient
medium with negative density (pressure) gradient develop a broad and
relaxed cocoon and the beam rapidly accelerates due to the drop of the
ambient density and ram pressure.  No internal knots are formed in
this case and the radiative shocks at the head are faint and will
provide little radiation (Fig.  1c). This is consistent with the
``invisible'' portions of the observed outflows, e.g., the gaps close
to the source of Haro 6-5B (FS Tau B) (Mundt, Ray and Raga 1991,
hereafter MRR), AS 353A (e.g., Hartigan, Mundt and Stocke 1986), and
1548 C27 (e.g., Mundt et al.  1987).  On the other hand, jets
propagating into portions of an ambient medium with increasing density
(pressure) have their cocoon/shroud compressed and pushed backwards by
the ambient ram pressure and the beam is highly collimated.  The
compressing medium promotes Kelvin-Helmholtz (K-H) instabilities which
cause beam pinching, twisting and flapping (Figs. 1a and 2). This
favors the formation of regularly-spaced traveling internal knots, and
an increase in the jet confinement, as required by observations, but
the bow shock-like structure at the head disappears.

Observed jets are, in general, well collimated and knotty and many
show one or more bright bow shock-like structures at the head (e.g.,
Reipurth 1989).  In GB94 (see also, e.g., Hartigan and Raymond 1992;
Raga and Kofman 1992; Stone and Norman 1993; Massaglia et al. 1995),
we showed that the internal traveling knots can be a product of rapid
variations in the jet ejection speed. We also have demonstrated that
bow shocks can be formed in jets with continuous ejection (GB93) or
with multiple outflow episodes of long period (GB94) and can keep
their structure approximately stable in portions of the jet where the
ambient medium is more homogeneous (GB93).  There are, however, some
observed jets whose visible structures resemble remarkably the
simulated jets of Fig. 2, which propagate into ambient gas with
increasing density gradient, as for example, Haro 6-5B-northern and HH
24G jets (e.g., Mundt et al. 1990, MRR).

Investigating a sample of 15 protostellar jets including those above,
MRR found that the flows are, in general, poorly collimated close to
the source (within distances $\sim \,$ few $10^{-4}$ pc) and become
highly collimated further out, on scales of the order of the jet
lengths ($\sim \, 10^{-2} - 10^{-1}$ pc), showing a decrease in the
jet opening angle with increasing distance. This clearly suggests that
a large scale collimation mechanism, probably due to the external
environment is at work (MRR). The Haro 6-5B and HH 24G jets, in
particular, narrow with increasing distance from the source, like the
jets of Figs. 2a and b.

The Haro 6-5B northern jet extends over $L_j \simeq 0.1$ pc and its
source is surrounded by an extended reflection nebula whose brightest
part lies on the jet axis at $ \sim \, 7 \times 10^{-3}$ pc. Within
$10^{-3}$ pc of the source, the jet has a strong lateral expansion
which is followed by a gap (between $ \sim 1$ and $1.7 \times 10^{-3}$
pc) where the jet is not seen. Further out, the jet is recollimated
and $narrows$ over most of its length. Its elongated structure is
knotty and $wiggling$.  After that, the jet becomes again invisible
and then shows up in a distant bow-shaped nebula ($ \simeq 0.1$ pc
from the source). As suggested by MRR, the initial expansion is likely
to be a signature of a highly overpressured and poorly collimated
flow. The appearance of the jet outside $1.7 \times 10^{-3}$~pc, where
the expansion is slow ($\Delta R / \Delta x \sim 0.025$), is
reminiscent of the jet brightening and collimation that we see in
simulations of cooling jets in an ambient medium of increasing density
(Figs.~2a and~b). K-H instabilities here would drive the formation of
internal shocks and small amplitude wiggles, so that the observed
bright knots might be oblique internal shocks excited by the K-H modes
(as seen in Fig. 2) instead of being due to intermittent jet ejection
(e.g., GB94). Outside this part of the flow, the disappearance of the
jet might arise from passage through an ambient medium with decreasing
density (and pressure), which would lead to a low emissivity (as in
Fig. 1c).  The distant bow-shaped nebula could mark an impact between
the jet and a last dense part of the ambient medium (e.g. GB93), or it
could be a relic of a much earlier outflow episode in a more
homogeneous ambient medium (e.g., GB94). The morphology of the Haro
6-5B counterjet is radically different from that of the jet,
suggesting that intrinsic asymmetries due to environmental effects are
at work (MRR).

Since we did not choose parameters for our simulations to fit the
particular case of the Haro 6-5B jet, our results can provide only
qualitative descriptions of the observations.  Nonetheless, we can
estimate the changes in the flow parameters that would be required to
produce the observed expansion and collimation of the Haro 6-5B jet.
The collimated part of the jet flow has 
and expansion rate $(\Delta R/ \Delta x)_{obs}
\simeq 0.025$ (MRR), and in our simulations of the $\eta = 3$ jet
propagating into an increasing pressure environment (Fig. 1a, Fig. 2b)
we find $\Delta R/\Delta x \sim 2\times$ smaller.  This suggests that a
reduction of the jet overpressure by a factor $\sim 4$
would cause a
better match between our simulations and the observed flow. We note
that the ambient temperature used in our simulations (and thus $p_a$)
is somewhat larger than the observed values in star forming regions,
an approximation that has been used in previous work (e.g., BFK, GB93,
GB94) partially to mimic the presence of a magnetic pressure. A
decrease of $T_a$ in our simulations would be consistent, for example,
with the required decrease of $p_a$ in order to fit the observed
expansion and narrowing of the Haro 6-5B jet.

Large amplitude side-to-side wiggles and knots are also observed in
HH30, HH83, HH84, HH85, HH110, and other jet systems (e.g., MRR,
Reipurth 1989).  Such features could also be due to the mechanism
discussed here.  We note, however, that some of these jets keep
broadening slowly with distance (e.g., HH30) or have an approximately
constant width over a section of their length (e.g., HH83).  This
would suggest less efficient confinement by the ambient medium in
these cases, perhaps because of a larger initial density ratio between
jet and ambient gas, as in the high $\eta$ jet of Fig. 2c, or a
smoother density stratification of the ambient medium (smaller
$\alpha$, see eq. 1).

Although, on average, the jets of Fig. 2 are decelerated by the
increasing ram pressure of the ambient medium, we find that the
velocity of advance of their working surfaces, $v_{ws}$, has an
oscillating pattern. Since the density ratio $\eta$ decreases with
distance, $v_{ws} \simeq \, v_j / [1 + (\eta \, \epsilon)^{-1/2} ]$
(see, e.g., eq. 1 of GB93) initially decreases as the jet propagates
downstream with a ratio of jet to head square radius $\epsilon \,
\equiv \, (R_j/R_h)^2 \, \simeq \, 1$.  Later, as the jet head is
compressed by the ambient medium, the increase of $\epsilon$ more than
compensates for the decrease of $\eta$, and the jet head accelerates
as it narrows. Eventually, the further decrease of $\eta$ dominates
the equation and the working surface decelerates again. For example,
the $\eta (x_o) = 3$ jet of Fig.  2b follows the $v_{ws} (\epsilon)$
relation above within $\sim \, 30 \% $. The internal shocks of Fig. 2
are regularly spaced (at $\sim \, $ 1 to 2 $R_j$) and travel
downstream with a similar variable velocity pattern. Their speed gets
closer to $v_{ws}$ as they approach the jet head.  Recent measurements
of the tangential velocity of the knots of Haro 6-5B (Eisl\"offel
1993) indicate an oscillating pattern in qualitative agreement with
the results above.  Further measurements of variations of the proper
motions of knots in systems like HH24G or HH83 would be an interesting
check on our predictions.  It is also important to note that the
variations of the shock velocity at the working surface imply
variations in the intensity of emission from the associated cooling
regions behind the shocks.

\acknowledgements
We are indebted to the referees (Colin Masson and an anomymous
referee) and to Tom Ray for their fruitful suggestions. Also EMGDP is
thankful to R. Mundt for his kind hospitality during the Workshop on
Disk and Outflows from Young Stars at the Max-Planck Institute in
Heidelberg and his relevant and fruitful comments on this work.  The
simulations were performed on HP Apollo 9000/720 and DEC 3000/600 AXP
Workstations whose purchase was made possible by the Brazilian Agency
FAPESP. This work has been partially supported by the Brazilian Agency
CNPq.

\newpage

\noindent
{\bf Figure Captions}

Figure 1. Central density contours of jets with $\eta (x_o) = 3$ and
$M_a = 24$. In a) the ambient medium has positive density gradient
($\beta = 5/3$, $\alpha = 0.5$), in b) it is homogeneous, and in c) it
has negative density gradient ($\beta =-5/3$, $\alpha = 0.5$). The
coordinates x and z are in units of $R_j$.  The contour lines are
separated by a factor of 1.3 and the peak of the density scale is $800
/ n_a(x_o)$.

Figure 	2. Jets with distinct $\eta(x_o)$ in an $\alpha = 0.5$ ambient
medium with positive stratification.  a) $\eta (x_o) =1$, b) $\eta
(x_o) = 3$, and c) $\eta (x_o) = 10$.  The initial conditions are the
same as in Fig. 1. The maximum density of the shells at the jet heads
are $n_{sh}/n_a(x_o) \simeq $211 (a), 638 (b), and 2540 (c).

\end{document}